\documentclass[twocolumn,english, showkeys]{revtex4-1}
\usepackage[utf8x]{inputenc}
\usepackage{textcomp}
\usepackage{graphicx}
\usepackage{hyperref}
\usepackage{amssymb}
\usepackage{latexsym}
\usepackage{babel}

\usepackage{fancyhdr}
\fancypagestyle{firstpage}{%
  \lhead{Manuscript accepted for publication in   \textit{Results in Physics}   (2019) 102233, final form published at \href{https://doi.org/10.1016/j.rinp.2019.102233}{https://doi.org/10.1016/j.rinp.2019.102233}.}

  \rhead{}
}

\begin{document}

\title{Laser slicing: a thin film lift-off method for GaN-on-GaN technology}%
\author{Vladislav Voronenkov$^{1,2}$ } \email{voronenkov@mail.ioffe.ru}

\author{Natalia Bochkareva$^{2}$ }

\author{Ruslan Gorbunov$^{1,2}$}

\author{Andrey Zubrilov$^{1,2}$}

\author{Viktor~Kogotkov$^{1}$}

\author{Philipp Latyshev$^{1}$}

\author{Yuri Lelikov$^{1,2}$}

\author{Andrey Leonidov$^{1}$}

\author{Yuri Shreter$^{1,2}$} \email{y.shreter@mail.ioffe.ru}

\affiliation{$^1$TRINITRI-Technology LLC,  St~Petersburg 194223, Russia \\ $^2$Ioffe Institute,  St~Petersburg 194021,  Russia}

\begin{abstract}
A femtosecond laser focused inside bulk GaN was used to slice a thin GaN film with an epitaxial device structure from a bulk GaN substrate. The demonstrated laser slicing lift-off process did not require any special release layers in the epitaxial structure. GaN film with a thickness of 5~$\mu$m and an InGaN LED epitaxial device structure was lifted off a GaN substrate and transferred onto a copper substrate.  The electroluminescence of the  LED chip after the laser slicing lift-off was demonstrated.
\end{abstract}
\maketitle

\thispagestyle{firstpage}


Bulk GaN substrate is a  crucial element in the production of reliable, high-current density GaN-based  devices.
 However, the widespread use of GaN substrates is currently limited due to their high cost. 
 Separation of the epitaxial  device structure from the bulk GaN substrate allows the reuse of the expensive  substrate multiple times and thus reduces the effective cost.
Also, the separation of the epitaxial structure is necessary to create heterogeneously integrated devices \cite{heterogeneous-gan-cmos-2009,HETEROGENEOUS-LED-2015,HETEROGENEOUS-2010},
as well as to improve  heat dissipation  \cite{DIAMOND-HEMT-2014-Kuball,HETEROGENEOUS-LED-2015} and light extraction \cite{LLO-InGaN-AKASAKI2014}. 

To lift-off GaN films from foreign substrates multiple methods were developed:
 laser lift-off \cite{LLO-Kelly-1996}; 
 natural stress-induced separation \cite{Natural-Stress-Moustakas2007,YAMANE20121};
  controlled spalling \cite{SPALLING-sapphire-LEDbedell-sadana-2013};
   chemical etching of ZnO \cite{CLO-ZnO-Rogers2007}, Ga\textsubscript{2}O\textsubscript{3}
\cite{CLO-TsaiGa2o3-2011}, CrN \cite{CLO-CrN-2008-Yao}, AlN \cite{CLO-AlN-2010-ChengFengLin}
and Nb\textsubscript{2}N \cite{nb2n-lift-off-2016} sacrificial layers;
doping-selective electrochemical etching \cite{CLO-nPlus-SiGaN-2009-Park};
substrate removal by grinding or etching \cite{melnik_nikolaev_nikitina_vassilevski_dmitriev_1997,ETCHING-GaAs-sumitomo-2001,Si-wafer-remove-Lesecq-2011,Mikulik-Si-etching-2013};
void-assisted separation \cite{USUI-VAS-2003};
 growth over patterned masks \cite{gogova2004elog,HENNIG2008911-WSiN-ELOG,lipski2010fabrication,AMILUSIK201399};
the use of weakly bonded release layers like graphene \cite{Graphene-IBM-kim2014principle},
BN \cite{BN-kobayashi2012layered}, carbon \cite{CARBON-Phil-2016};
and the use of substrates with easy cleavage along the c-plane of a GaN epilayer
\cite{Ga2O3-lift-off-Gogova2012,SCAM-Matsuoka-2017}. 

To separate a GaN film from a native GaN substrate several approaches were proposed:
 chemical lift-off process \cite{InGaN-GaN-PEC-Youtsey2017};
porous release layers created by chemical \cite{LEE-KOH-POROUS-ETCH}, electrochemical \cite{Mynbaeva-porous-liftoff,porous-lif-off-2013} or dry \cite{H2-porous-etching-YEH2011}  etching of GaN substrate;
 controlled spalling \cite{SPALLING-bedell-sadana2017};
  laser lift-off with an InGaN release layer \cite{LLO-InGaN-AKASAKI2014};
   ion implantation \cite{Smart-Cut-GaN-Tauzin2005,Moutanabbir2010-ion-cleaving,Ion-Cut-Samsung-2013,huang2018defects};
and free-carrier-absorption laser lift-off \cite{FCA-LLO-Virko2016}. 

The lift-off methods based on the use of an intermediate layer require
a complicated epitaxy process. The ion implantation method does not
allow separating epitaxial device structures due to a large number
of point defects formed by the implantation process \cite{Ion-Cut-Samsung-2013,He-ion-damage-HAN2017,huang2018defects}.
The free-carrier-absorption LLO method is limited to separating undoped films from a heavily doped substrate \cite{FCA-LLO-Virko2016}.

We have proposed a laser slicing lift-off (LSLO) method for separating an epitaxial structure
 from a bulk GaN substrate, one that does not require any special release layers and  can be used both for lifting off pure GaN films and for lifting off GaN films with device structure \cite{shreter2013method}. The method is based on the effect of GaN decomposition induced by an ultra-short laser pulse focused inside a bulk GaN material. The use of femtosecond lasers is preferable as it allows to reduce the heat-affected zone and the corresponding damage in the surrounding material \cite{stuart1996-ns-to-fs}.
This effect  was earlier used to create hollow 3D-channels inside bulk GaN \cite{FS-GaN-MACHINING-MIDORIKAWA-2009}
and to dice transparent wafers with laser stealth dicing technology~\cite{LSD-Kumagai-2007}. 

\begin{figure}[!t]
\centering
\includegraphics[width=\linewidth]{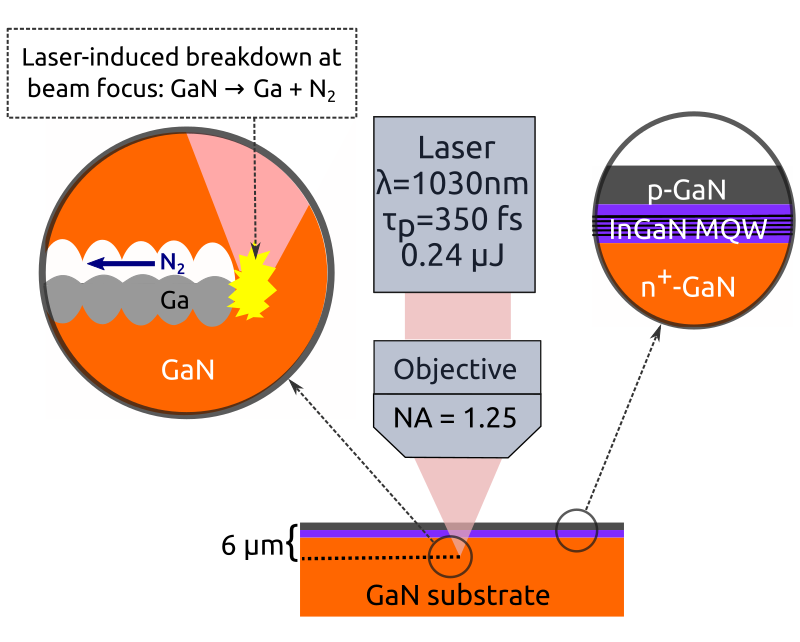}
\caption{Experimental setup for the laser slicing 
lift-off process: a beam of Yb-doped femtosecond laser with a pulse width
of 350~fs and a pulse energy of 0.24~$\mu$J is focused inside a GaN wafer, 6~$\mu$m below the surface. The decomposition of GaN takes place in the focus area of the laser beam. The wafer is scanned in the $X$-$Y$ plane until a continuous layer of decomposed material is created inside the wafer.}
\label{fig:FS-LO-SETUP}

\end{figure}

The principle of the LSLO method is as follows.
To lift-off a thin GaN film with an InGaN multiple quantum well (MQW) device structure from a bulk GaN substrate, the pulses of a near-infrared femtosecond laser are focused  inside the GaN layer, several microns under the surface (fig.\ref{fig:FS-LO-SETUP}). The laser photon energy is less than the InGaN bandgap, so both the InGaN device structure and the GaN substrate are transparent at the wavelength of the laser. The laser pulse energy is adjusted so that the nonlinear breakdown threshold is exceeded only in the focus area and the GaN material in this area is decomposed. Scanning of the focus position  is performed in the $X$-$Y$-plane until a continuous layer of decomposed material is created inside the wafer.

\begin{figure}[!t]
\centering
\includegraphics[width=0.8\linewidth]{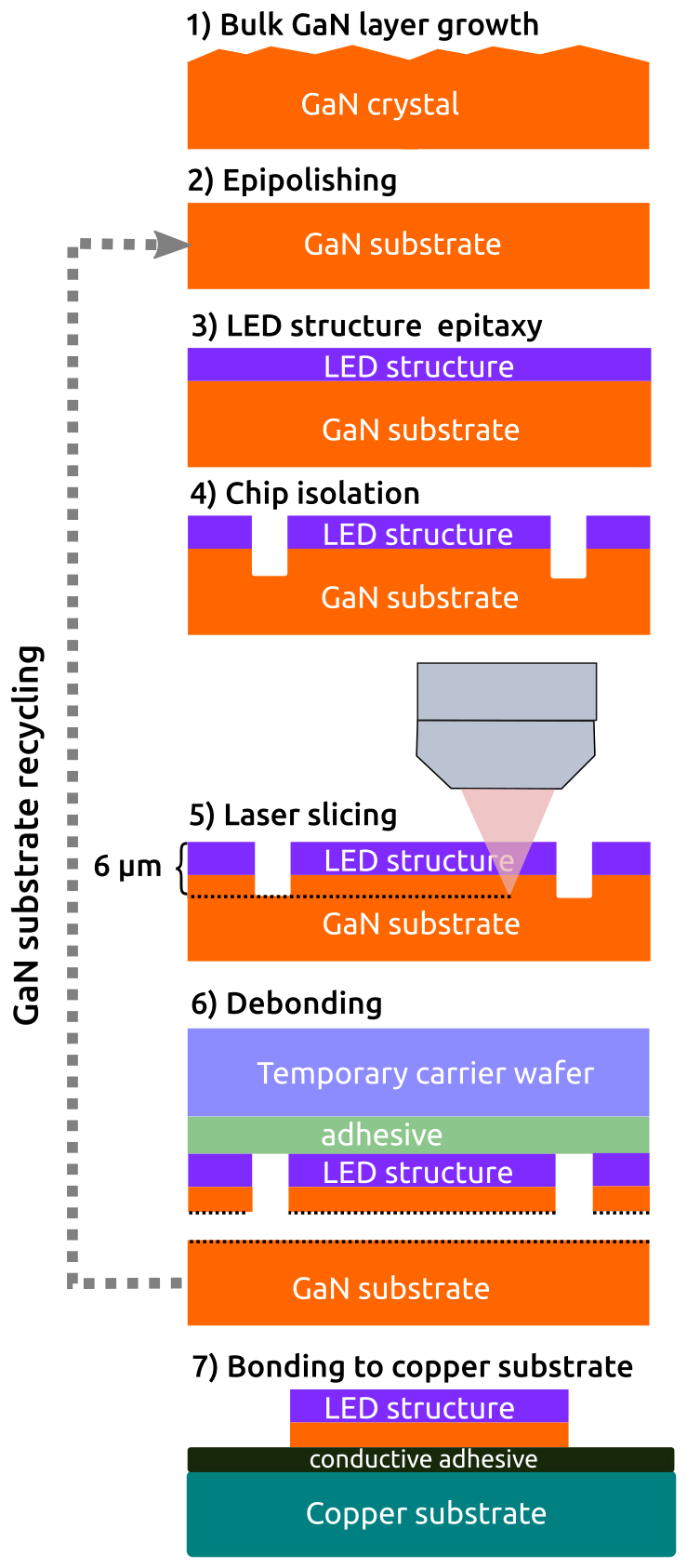}
\caption{Schematic diagram of thin-film InGaN LED chip fabrication using the laser slicing lift-off process. }
\label{fig:Schematic-diagram-of}

\end{figure}

In this work, a proof-of-concept LED chip  fabrication using the LSLO method is described. First, a 5-$\mu$m thick GaN film with an InGaN LED structure was grown homoepitaxially on a bulk GaN layer. After that, the film with the LED structure was lifted-off from the substrate and transferred to a copper carrier (fig.~\ref{fig:Schematic-diagram-of}).  The electroluminescence of the  LED chip after the lift-off was demonstrated. A detailed study of the LED chip for possible damage  caused by the LSLO process will be published separately.

A GaN-on-sapphire HVPE template was employed as a bulk-like
substrate for the InGaN LED structure epitaxy. The template was grown on a 2-inch c-plane (0001) sapphire substrate using a two-stage growth process \cite{voronenkov2013two}. The thickness of the bulk-like HVPE GaN layer was 50~$\mu$m. The surface of the HVPE GaN template was epi-polished by a chemical-mechanical polishing process. After that, the InGaN MQW LED structure was deposited using a metal-organic chemical vapor deposition system. The LED structure consisted of a 5-$\mu$m thick $n^{+}$ doped GaN:Si ($n$=$4\cdot10^{18}$~cm$^{-3}$) layer; five pairs of InGaN quantum wells with a well width of 2.3~nm and a barrier width of 11~nm; and a 0.3-$\mu$m thick p-doped GaN:Mg layer (p=$3\cdot10^{17}$~cm$^{-3}$). 

The chip isolation and the LSLO process were performed with an ytterbium-doped, diode-pumped solid-state laser with a wavelength of 1030~nm, a pulse duration of 350~fs, and a pulse energy of 400~$\mu$J.  A three-axis  motorized translation  stage was used  to perform scanning in the $X$-$Y$-plane and to adjust the focus position along the $Z$ axis.

First, a 500$\mu$m$\times$500$\mu$m  chip was isolated through the laser scribing process -- grooves with a depth of 6~$\mu$m and a width of 5~$\mu$m were cut along the chip perimeter. Such grooves are necessary to release the gas produced by the GaN decomposition during the LSLO process. A 20$\times$ dry objective was used to focus the laser beam on the surface of the GaN layer. The pulse energy was 5~$\mu$J, pulse repetition rate was 1~kHz and the scanning velocity was 1000~$\mu$m/s. 

To perform the LSLO an oil immersion objective (100$\times$, NA = 1.25) with an $n$=$1.518$ immersion oil was used. Before performing the LSLO, the breakdown threshold was determined: the laser was focused at a depth of 6 $\mu$m below the surface and the laser pulse energy was gradually increased until a breakdown in the focus area was observed at a pulse energy of 0.2 $\mu$J. 
After that, the LSLO was performed using a pulse energy of 0.24~$\mu$J. The laser beam, directed from the front side of the sample with a beam focus located 6~$\mu$m under the surface was used to create a thin decomposed layer 6~$\mu$m below the surface of the wafer. The scanning pattern exhibited a set of parallel rows. The scanning velocity along the row was 1000~$\mu$m/s and the  pulse repetition rate was 2.5~kHz resulting in a pulse-to-pulse distance along the row of 0.4~$\mu$m; the distance between the rows was 2~$\mu$m.

\begin{figure}[!t]
\centering
\includegraphics[width=\linewidth]{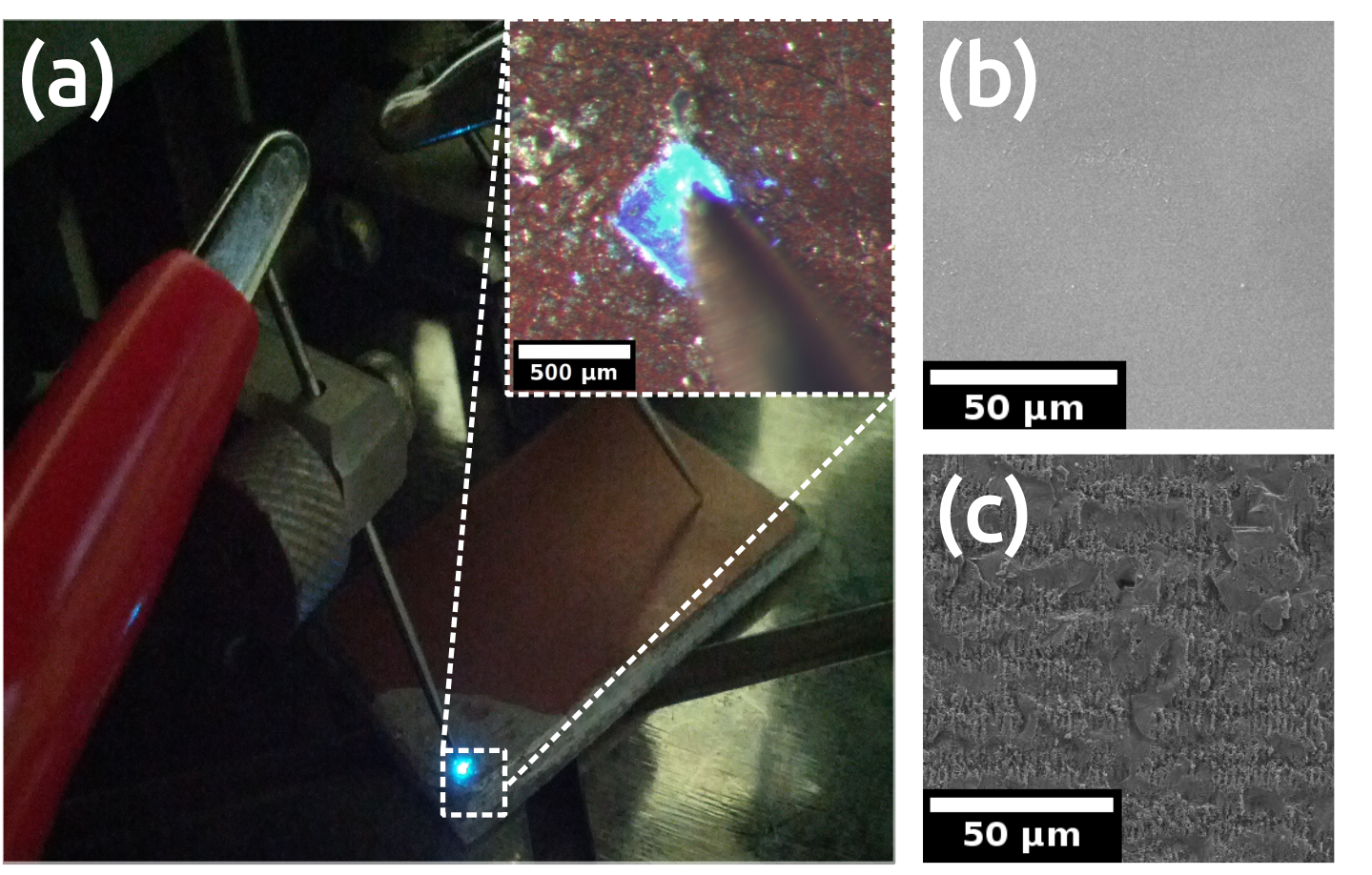}
\caption{a) A photograph of the InGaN LED chip on a copper substrate with a current probe connected, electroluminescence at 1~mA operating current. The inset shows a microphotograph of this chip. b) SEM image of the Ga-face of the chip after the LSLO process. c)  SEM image of the N-face of the chip after the LSLO process.}
\label{fig:RESULT}

\end{figure}

After the LSLO process,  the LED chip was bonded to a temporary carrier wafer using a 
cyanoacrylate adhesive. After the adhesive  hardened, the chip was debonded from the substrate. Then, the cyanoacrylate adhesive was dissolved, and the chip was bonded to a copper substrate using an electrically conductive silver epoxy as a bonding material. 

To observe the electroluminescence of the chip, an electrical current probe was connected directly to the surface of the chip, without applying any current-spreading layers.
The light emitting image of the  LED chip on a
copper substrate is shown in fig. \ref{fig:RESULT}(a). The
operating current was 1~mA. Blue-light emission was observed under
the probe, confirming that the MQW LED structure was not destroyed during the LSLO process.
A detailed study of the I-V characteristics and the electroluminescence spectrum, as well as a detailed  comparison of the LED performance  before and after the LSLO process will be published separately.

The SEM image of the top side (Ga-face) of the LED chip after the
LSLO process  is shown in fig.~\ref{fig:RESULT}(b). No signs of damage caused by the LSLO were observed on the surface.
The SEM image of the bottom side (nitrogen-face) of the LED chip  after  LSLO and debonding is
shown in fig.~\ref{fig:RESULT}(c). The remains of the decomposed layer were observed. 
The surface roughness of the bottom side of the chip was estimated using a tilted-view SEM image (not shown) to be about 1$~\mu$m, which corresponds to the  focus depth of the laser beam. 
The thickness of the separated film was measured using a side-view SEM image (not shown) and was 5~$\mu$m, which corresponds to the position of the focus below the surface.

Following is a brief discussion of  the technological limits and possible improvements of the method.
The maximum thickness of the lifted-off layer is limited by the defocusing of the laser beam caused by spherical aberration and optical inhomogeneity of the material. Recent advances in the growth of pure GaN \cite{SCIOCS-2017-pure} make it possible to obtain an optically homogeneous material,  and the adaptive optics techniques can compensate for spherical aberration when focusing to a depth of hundreds of microns \cite{Salter-2014-depth}.
The minimum thickness of the decomposed layer is limited by the depth of focus of the laser beam, which can be reduced by the use of high-numerical-aperture objectives, such as  solid immersion lenses \cite{SIL-Kino-1999}. 
Processing throughput is mainly limited by the scanning speed,  therefore the use of  parallel multi-beam machining methods \cite{Kato-2005-microlens, Booth-2010-parallel}   is highly desirable.
Wafers with already deposited  metallization layers can also be processed. For this,  the focused laser beam is directed from the underside of the wafer.
 Other materials that decompose irreversibly under the action of laser breakdown can also be processed. The greatest practical interest of such materials is diamond, which is graphitized by laser breakdown \cite{Konov-2008-diamond}.

In conclusion, a 5-$\mu$m thick GaN film with the InGaN LED MQW structure
was lifted off from the bulk GaN layer by focusing near-infrared femtosecond laser pulses
6~$\mu$m under the surface of the GaN wafer. The film was transferred to
a copper carrier and the electroluminescence of the LED structure was demonstrated.

\begin{acknowledgments}
The authors gratefully thank Dr. I.~Guiney and Prof. C.~Humphreys
from Cambridge University for the InGaN LED structure growth.
\end{acknowledgments}
\bibliographystyle{IEEEtran}
\bibliography{FS-LED-LO}

\end{document}